\documentclass{DISproc}
\usepackage{epsfig}
\begin{document}
\title{Summary of the Structure Functions Working group at DIS 2012}

\author{{\slshape Amanda Cooper-Sarkar$^1$, Pedro Jimenez-Delgado$^2$, 
Ringail\.e Pla\v cakyt\.e$^3$, 
}\\[1ex]
$^1$Denys Wilkinson Bdg, Keble Rd, OX1 3RH OXFORD, UK \\
$^2$Institut f\"ur Theoretische Physik, Universit\"at Z\"urich, 8057 Z\"urich, Switzerland\\
Jefferson Lab, 12000 Jefferson Avenue, Newport News, VA 23606, USA\\
$^3$DESY, Notkestra{\ss}e 85, 22607 Hamburg, Germany \\}
\contribID{xy}

\doi  

\maketitle

\begin{abstract}
  A summary of the recent experimental, phenomenological and 
theoretical results presented 
 in the Structure Functions working group at DIS2012 workshop. 

\end{abstract}

\section{Introduction}
 In the Structure Functions working group experimental results relevant to 
the determination of parton distributions were presented by H1, ZEUS, ATLAS, 
CMS and LHCb. The HERAfitter tool, which is an open 
access code for fitting PDFs to relevant data, and an
 update of the FastNLO tool, which allows the fast computation 
of higher-order cross sections at hadron colliders, were presented. 
 Progress in the determination of the parton distributions of the nucleon from 
most global PDF groups was reviewed and more restricted studies which focus 
on particular aspects of these determinations were also discussed. 
Analyses of nuclear PDFs were presented. Finally, phenomenological 
contributions within frameworks which appear as extensions or alternatives 
to the usual collinear factorization approach were discussed.

\section{Summary of the presentations}

\begin{figure}[htb]
 \begin{minipage}[b]{0.5\linewidth} 
   \centering
  \includegraphics[width=0.9\textwidth]{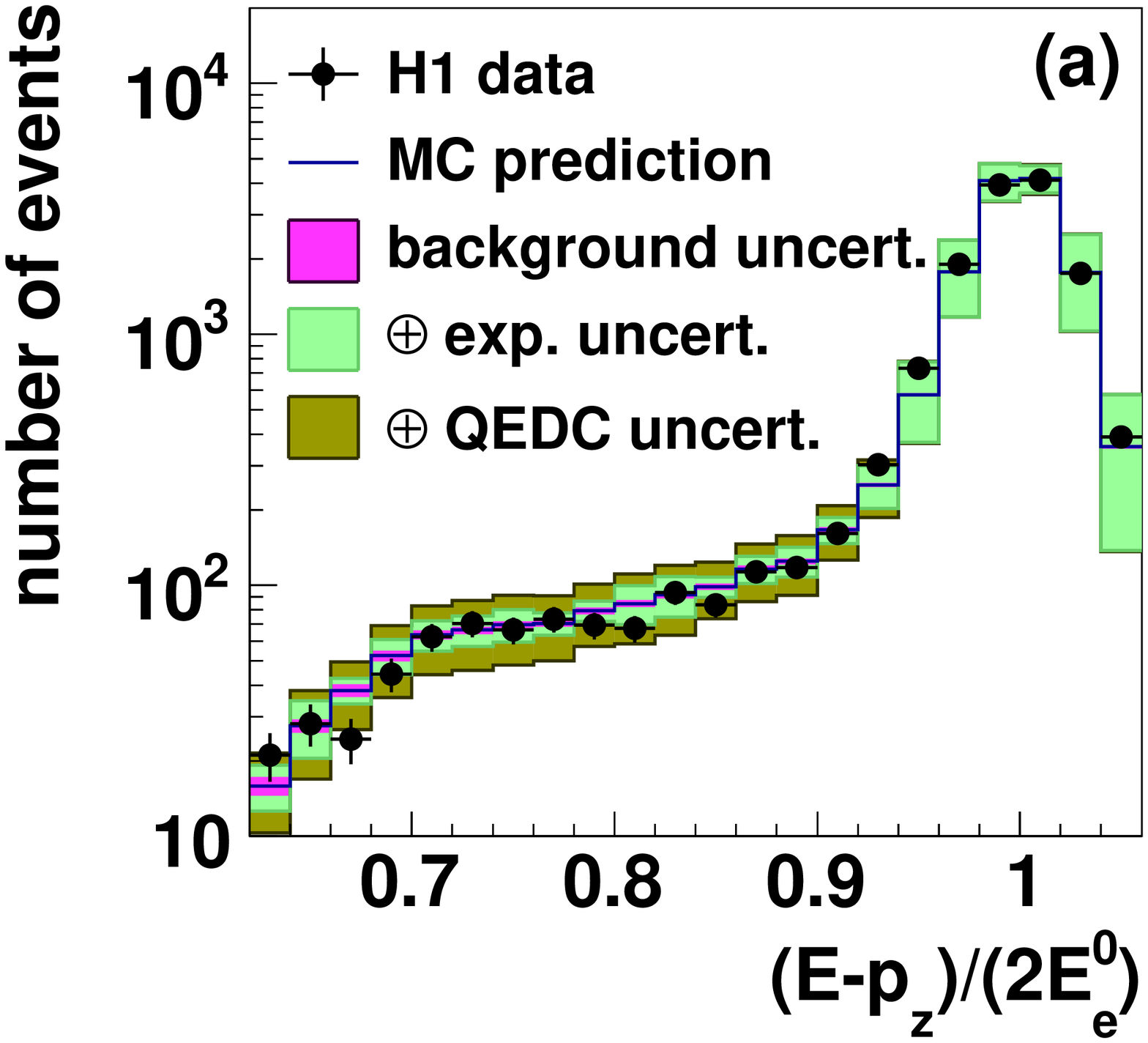}
 \end{minipage}
 \begin{minipage}[b]{0.5\linewidth} 
   \centering
  \includegraphics[width=0.9\textwidth]{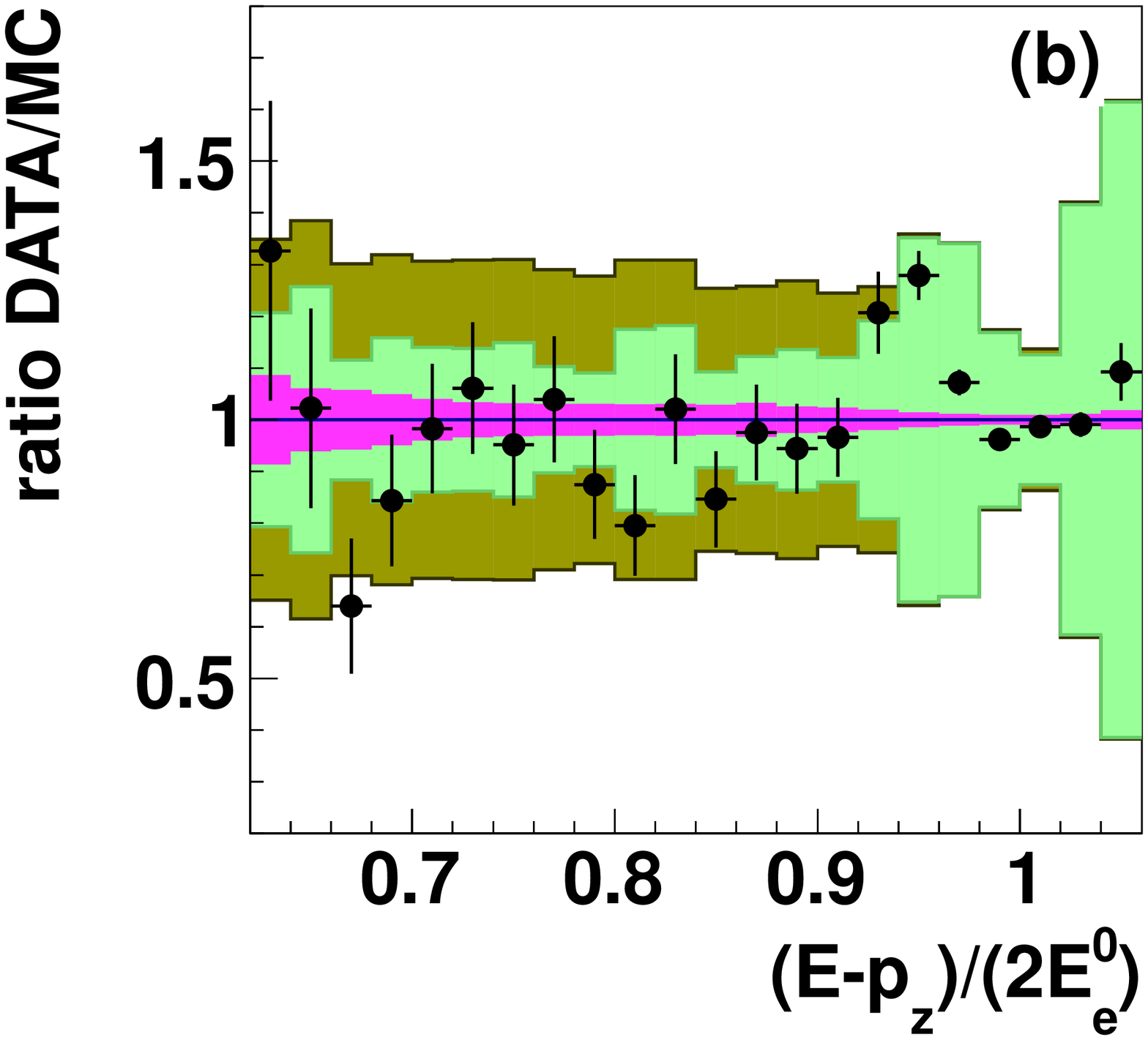}
 \end{minipage}
  \caption{Distribution of the variable $(E − p_z)/(2E_e^0)$ (see text) calculated from the sum of the electron
          and photon four-momenta. In (a) the event counts are shown and in (b) the ratio of data
          to expectation is drawn. The data are shown as black dots with the statistical uncertainties
          indicated as vertical bars and the simulation (including background) is indicated as a solid line, 
          with various components of the systematic uncertainty shown as shaded areas.}
  \label{Fig:h1lumifig}
\end{figure}

A measurement of the integrated luminosity of HERA data collected in the years 2003 to 
2007 and based on the elastic QED Compton process $ep \to e\gamma p$ has been performed 
by the H1 Collaboration~\cite{qedcompton}. 
Contrary to the standard HERA luminosity measurement which exploits
Bethe-Heitler (BH) scattering with electron and the photon emitted almost collinearly 
to the incident electron, the particles in QED Compton scattering have a sizable 
transverse momentum with respect to the incident electron and are detectable in 
the main H1 detector. The advantage of this method is its insensitivity 
to the details of the beam optics, but its disadvantage is limited statistical precision.
The precision of the experimental and theory uncertainties in this analysis
 are illustrated using the variable $(E - p_z)/(2E_e^0)$ in 
Fig.~\ref{Fig:h1lumifig}. This variable is calculated from the sum of the 
four-momenta of the electron and the photon 
(where $E_e^0$ is electron beam energy) and is expected to peak at unity. 
The tail to small values of $(E − p_z)/(2E_e^0)$ originates from initial state radiation
(where theory uncertainty dominates), whereas values larger than unity occur due to 
resolution effects. 
The measured integrated luminosity is determined with a precision of $2.3\%$ and is 
in agreement with the Bethe-Heitler measurement which has larger uncertainty for the 
second period of HERA data taking.

ZEUS have completed the measurement of inclusive cross sections from 
HERA-II running by finalising the Neutral Current (NC) 
$e^+p$ measurement~\cite{frederike}. 
The measurements are based on an integrated luminosity of 
$135.5$ pb$^{-1}$ taken in 2006 and 2007 at a centre-of-mass energy of $318$ GeV.
The double-differential cross sections in $Q^2$ and $x$ have been
measured in the kinematic region $Q^2 > 185~$GeV$^2$ for both positively and 
negatively polarised positron beams. 
These measurements have been used to extract the polarisation asymmetry 
parameter
$
A^+ = \frac{2}{(P_+ - P_-)} \frac{(\sigma^+(P_+) - \sigma^+(P_-))}{(\sigma^+(P_+) + \sigma^+(P_-))},
$
where $P_+ = +0.32$ and $P_-= -0.36$ denote the magnitude of the beam 
polarisations and $\sigma^+(P)$ denotes the cross section measured at 
polarisation $P$. This quantity is sensitive to the electro-weak vector couplings 
of the quarks and the non-zero asymmetry observed 
is a direct measure of parity 
violation. 
\begin{figure}[htb]
  \centering
\centerline{
\epsfig{figure=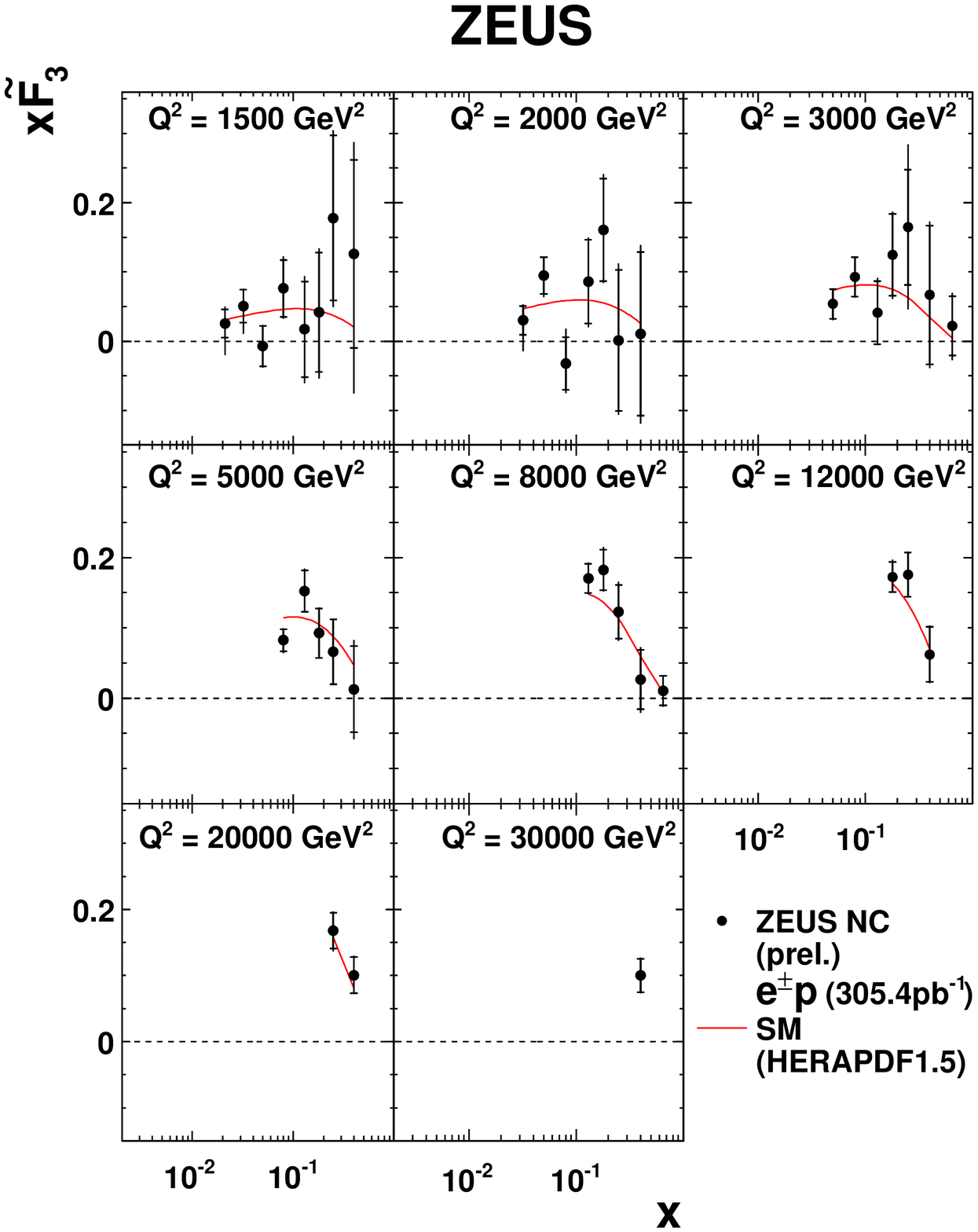,width=0.45\textwidth}
\epsfig{figure=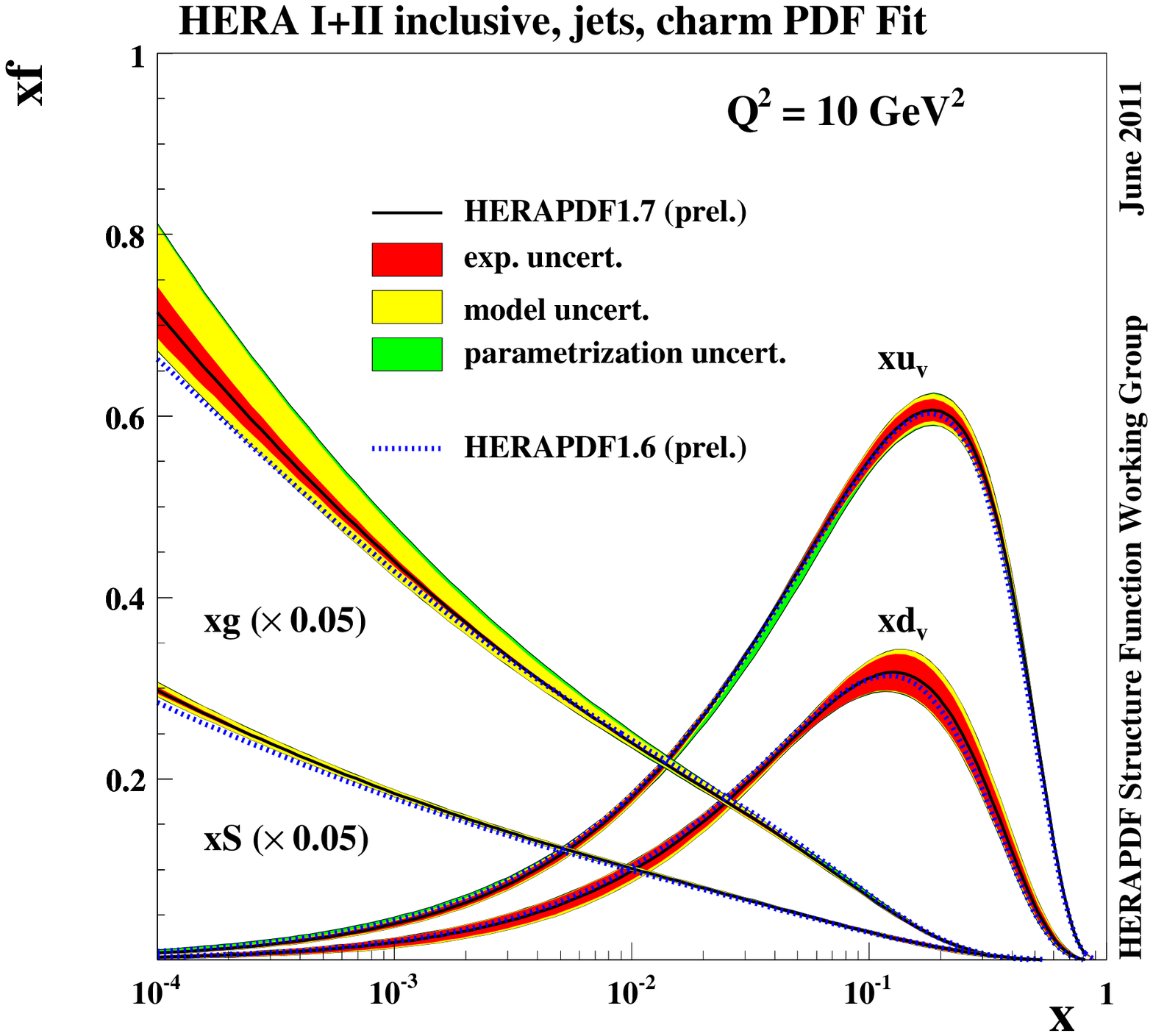,width=0.45\textwidth}}
  \caption{
Left: the structure function $xF_3$ as a function of $x$ in $Q^2$ bins, 
compared to the predictions of HERAPDF1.5.
Right: Parton distribution functions for HERAPDF1.7 }
  \label{fig:xf3herapdf17}
\end{figure}
The NC $e^+p$ data may also be averaged over positive and negative 
polarisations and 
combined with $169.9~$pb$^{-1}$ of NC $e^-p$ measurements~\cite{ncem} to yield 
the structure function $xF_3$, which gives information on valence parton 
distribution functions. The measurement of $xF_3$ as a function of $x$ in 
$Q^2$ bins is shown in Fig.~\ref{fig:xf3herapdf17}, 
compared to predictions from HERAPDF1.5

The H1 and ZEUS experiments have already combined their data from HERA-I 
running~\cite{heracomb} and have made preliminary combinations of HERA-II data 
from nominal energy and low energy running. A preliminary combination of 
$F_2^{c \bar{c}}$ data has also been made. These combined data sets, 
together with data on inclusive jet production from both H1 and ZEUS, 
have been used as the input to extract parton distribution functions (PDFs) 
in the HERAPDF1.7 NLO QCD fit~\cite{krysz}. 
These PDFs are illustrated in Fig.~\ref{fig:xf3herapdf17}.
All of the input data sets are well fit and consistent.
In comparison to the published HERAPDF1.0 PDFs~\cite{heracomb}, 
which were based only on the 
HERA-I combination, the HERA-II high $Q^2$ cross-sections further constrain
 the high-$x$ valence PDFs; the HERA-II low energy runs help to constrain the 
low-$x$ gluon PDF; the charm data constrain heavy quark schemes and the jet 
data help to constrain $\alpha_s(M_Z)$.

\begin{figure}[htb]
 \begin{minipage}[b]{0.59\linewidth} 
   \centering
  \includegraphics[width=0.74\textwidth]{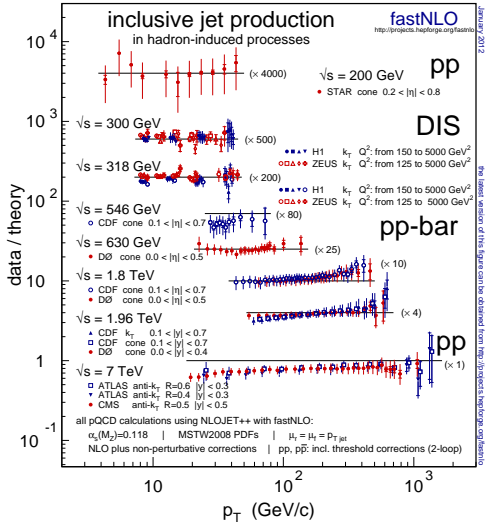}
 \hspace{0.5cm} 
 \end{minipage}
 \begin{minipage}[b]{0.39\linewidth}
  \caption{An overview of data over theory ratios for inclusive jet cross sections,
          measured in different processes at different center-of-mass energies. The data are
          compared to calculations obtained by FastNLO. The inner error bars represent
          the statistical errors and the outer error bars correspond to the quadratic sum of
          all experimental uncertainties.}
  \label{Fig:fastNLOfig}
 \end{minipage}
\end{figure}

Using jet production data in PDF fitting 
requires fast repeated computation of NLO 
jet cross sections. The  
FastNLO~\cite{fastNLOpaper} package provides a method to store the matrix 
elements calculated for such
higher-order cross-sections on grids such that the cross-sections may be 
calculated quickly by convolution of these grids with the input PDFs. 
The package can be used for jet cross-sections from DIS and from 
hadron colliders. Typical applications of FastNLO are data 
and theory comparisons for various PDF sets, derivation of scale uncertainties, determination of $\alpha_s$.
Fig~\ref{Fig:fastNLOfig} shows the comparison of the inclusive jet data
from various experiments 
to theory predictions obtained with FastNLO. 
Version 2 of the FastNLO project offers variaty of new features for users including
largely improved technical aspects of the code (like improved reading tools) and
flexibility in e.g. scale composition or scale variation. More details about new features in 
the version~2 of the FastNLO can be found in~\cite{fastNLOtalk}. 

The HERAFitter tool has been develpoed by H1 and ZEUS as an open access code 
for fitting PDFs to relevant data~\cite{voica}. 
Whereas the HERAFitter has been developed 
from the HERAPDF QCD fitting framework, based on QCDNUM for the NLO and NNLO 
QCD evolution, it goes far beyond this. The package 
can be used to fit all types of data used in a global PDF fit: inclusive cross 
sections, heavy quark structure functions, jet production data from Deep 
Inelastic Scattering (DIS); inclusive cross-section from fixed target data; 
Drell-Yan (including $W,Z$) cross sections from fixed target data and from 
Tevatron and LHC data; jet cross-sections from Tevatron and LHC data. 
A variety of options 
which facilitate benchmarking are available with the package: experimental 
systematic uncertainties can be treated as correlated or uncorrelated and PDF 
uncertainties may be evaluated from the Hessian covariance matrix or by the 
generation of Monte-Carlo replicas; the structure functions may be 
computed in a variety of heavy quark schemes; the package is interfaced 
to both FastNLO and to Applgrid for the fast and correct input of NLO jet 
cross sections and Drell-Yan cross sections and the input of top cross 
sections via HATHOR is under development. Plotting tools are provided for the
fit output and the resulting PDFs are supplied in the LHAPDF format.
The package is under continuous development to provide a common platform for 
useful tools; for example the NNPDF reweighting tool has been 
included to allow fast computation of the impact of new data on existing PDFs,
and the package has recently been extended to
make fits to diffractive cross sections and to make fits using dipole models.

\begin{figure}[htb]
 \begin{minipage}[b]{0.5\linewidth} 
   \centering
  \includegraphics[width=0.82\textwidth, angle=270]{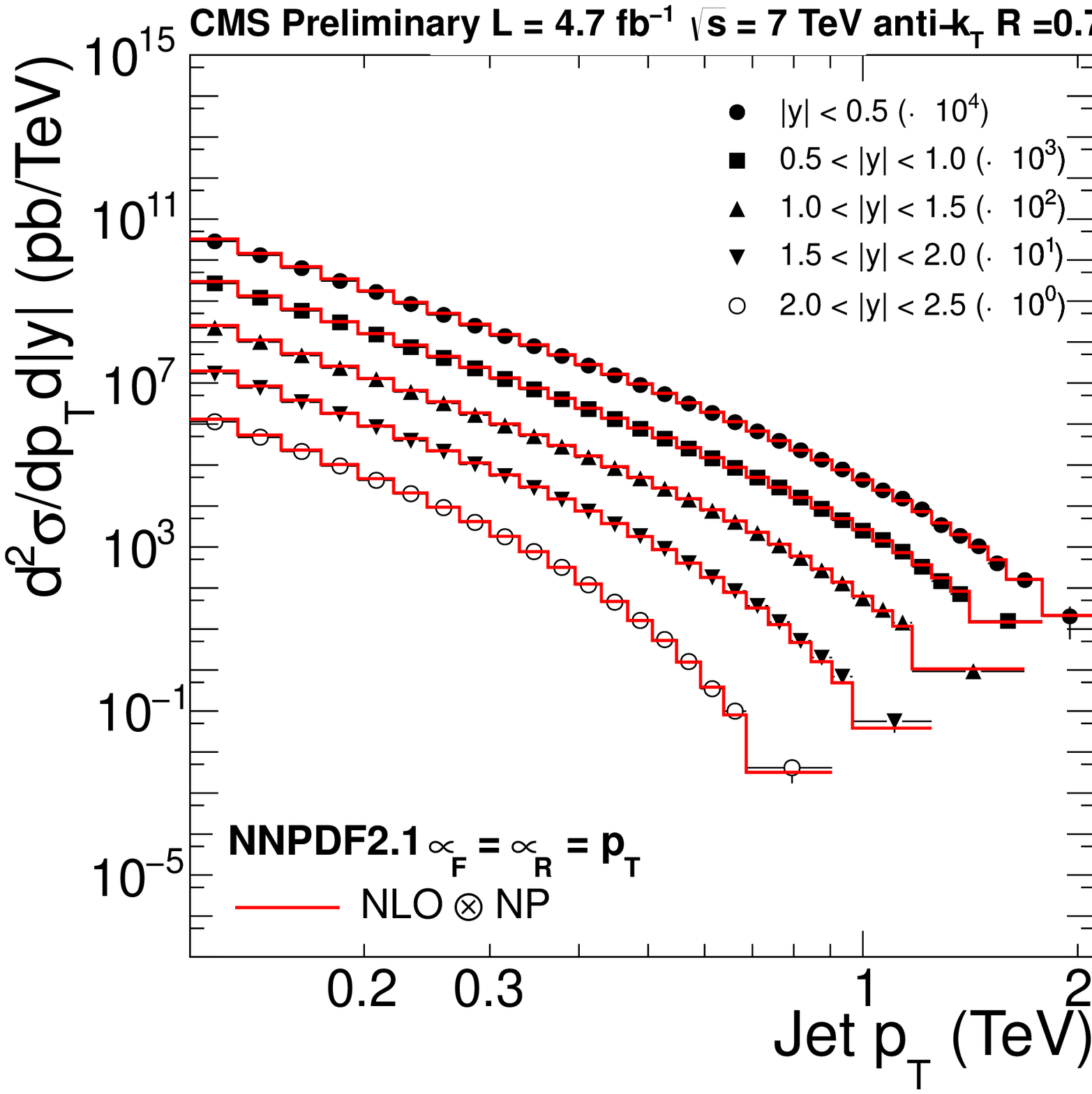}
 \end{minipage}
 \begin{minipage}[b]{0.5\linewidth} 
   \centering
  \includegraphics[width=0.72\textwidth, angle=270]{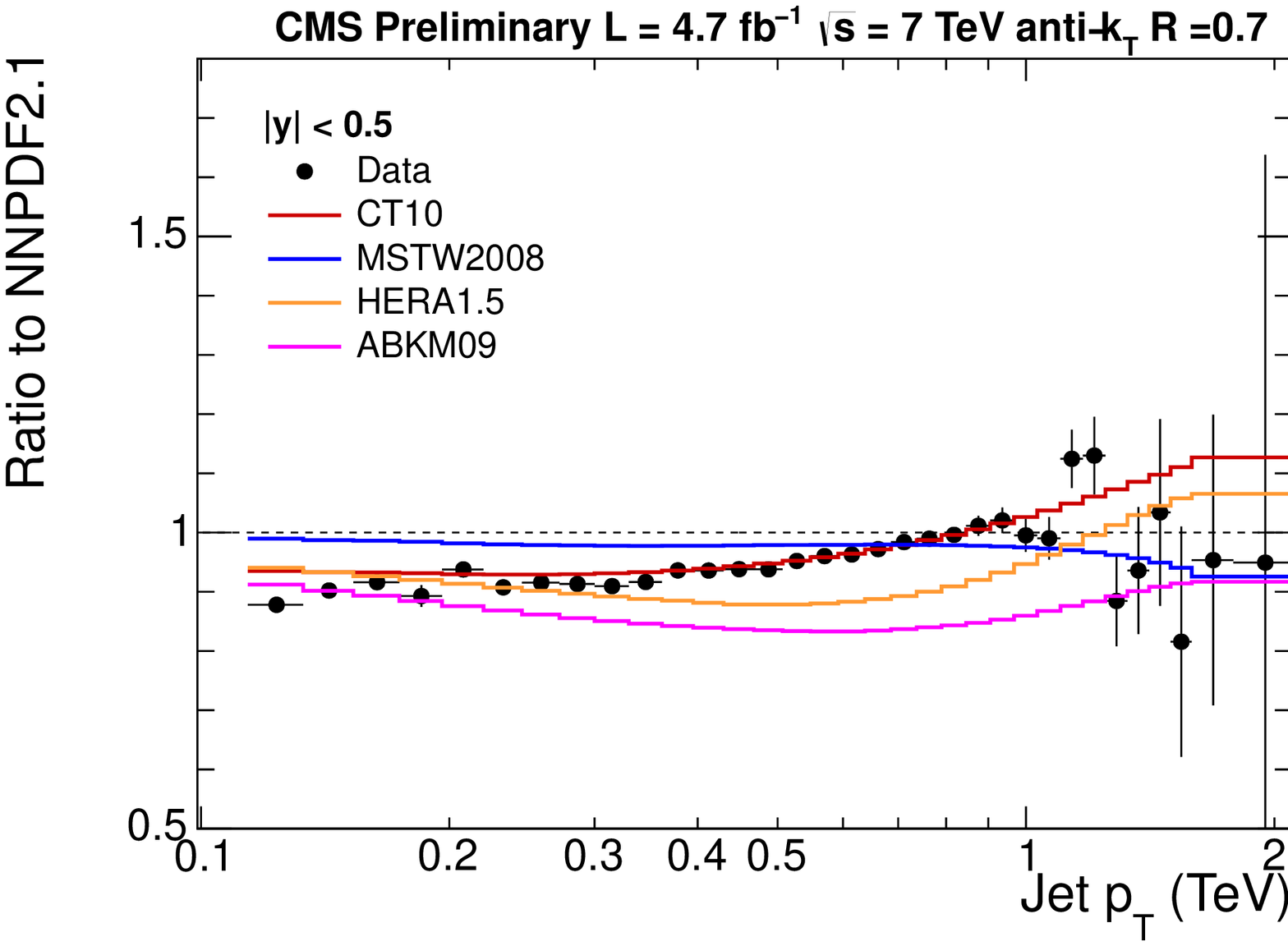}
 \end{minipage}
  \caption{ Inclusive jet cross sections (left) and ratio of inclusive jet (right) measured at CMS 
      compared to theoretical prediction using the central value of the NNPDF PDF set. 
     The solid histograms in the ratio plot also show expectations obtained with other PDF sets.}
  \label{Fig:jetcmsfig}
\end{figure}

CMS has measured differential jet cross sections using $\sqrt s = 7~$TeV data corresponding 
to $4.6$ pb$^{-1}$ of 2010 data~\cite{CMSjetTalk}. 
Reconstructed jets in this measurement cover rapidity up to~$|y|~=~2.5$, 
transverse momentum up to $p_{T} = 2~$TeV and dijet invariant mass up to $M_{JJ} =
5~$TeV.
The measured cross sections are compared to perturbative QCD predictions at next-to-leading order
using various sets of PDFs. Fig~\ref{Fig:jetcmsfig} illustrates the inclusive jet cross sections
together with theoretical predictions obtained using the central value of the NNPDF set (left) and
ratio of these cross sections to the NNPDF prediction with predictions from other PDF sets also shown (right).
Experimental and theoretical uncertainties in the measurement are comparable in size so that these data should be able to constrain PDF uncertainties.  
The systematic uncertainty correlations which are necessary for the PDF fits are in 
preparation.

\begin{figure}[htb]
  \centering
\centerline{
\epsfig{figure=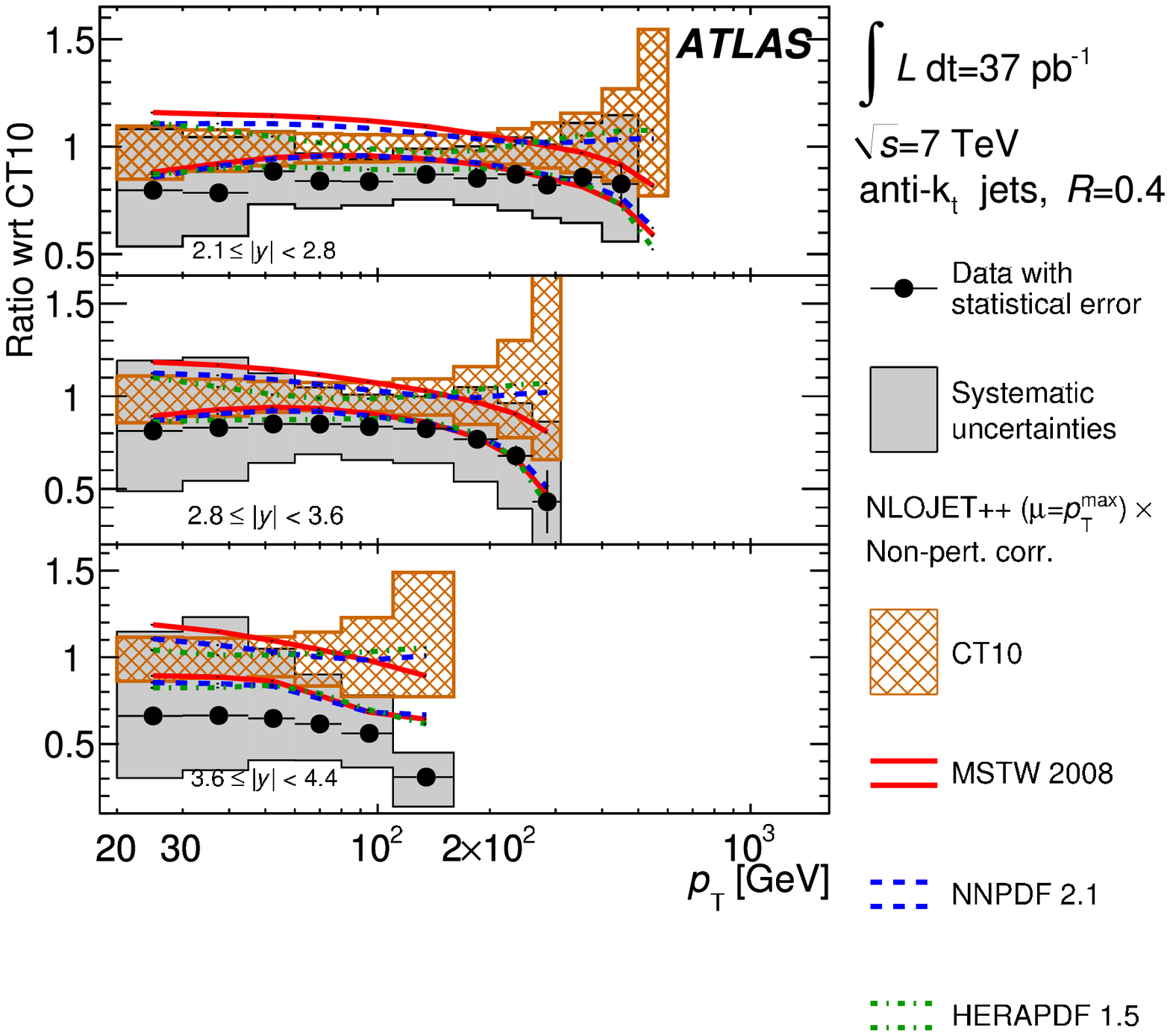,width=0.45\textwidth}
\epsfig{figure=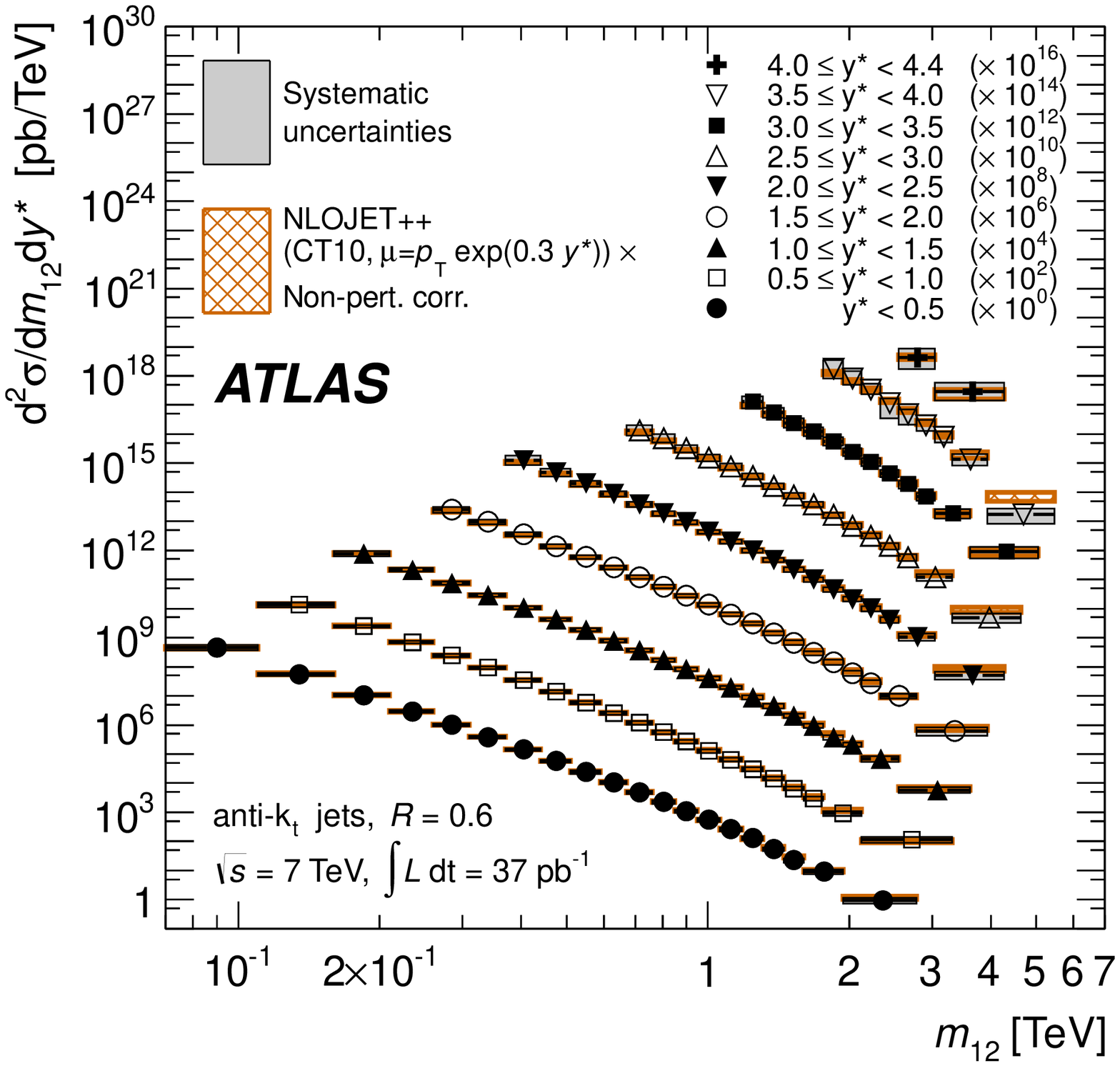,width=0.45\textwidth}}
  \caption{Left: ATLAS data on inclusive jet cross sections as a function of 
$p_t$ in high rapidity bins. The data are shown in ratio to the predictions of 
NLOjet++ made using CT10 PDFs, with other PDF predictions shown for comparison.
Right: ATLAS di-jet data as a function of di-jet invariant mass in rapidity 
bins}
  \label{fig:atlasjets}
\end{figure}

ATLAS has made a measurement of inclusive jet production using $36~$pb$^{-1}$ of 
data from 2010 running~\cite{Malaescu}. The data are provided with full information on 
correlated systematic uncertainties and this allows them to have some 
constraining power on PDFs. These data are well fit by most modern PDF sets 
such as MSTW08, CT10, NNPDF2.1 and HERAPDF1.5, however they prefer a somewhat 
less hard high-$x$ gluon than the Tevatron jet data. Fig.~\ref{fig:atlasjets} 
shows a comparison of these data to current PDFs. 
Dijet data from $4.7$fb$^{-1}$ of 2011 data are also available~\cite{Malaescu}
extending the range of the di-jet mass to $4.8~$TeV. These data are also 
illustrated in Fig.~\ref{fig:atlasjets}.

\begin{figure}[htb]
  \centering
\centerline{
\epsfig{figure=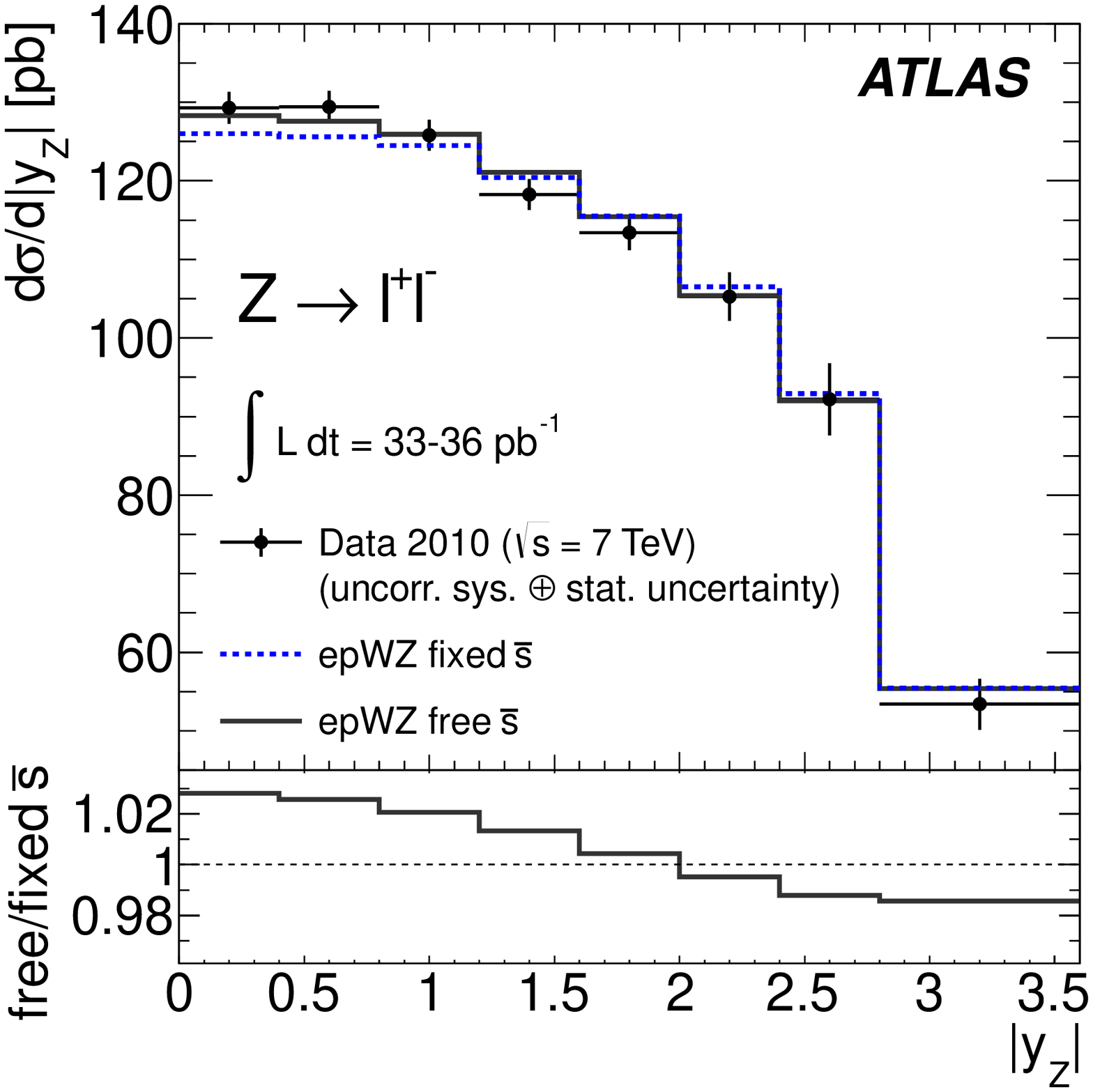,width=0.45\textwidth}
\epsfig{figure=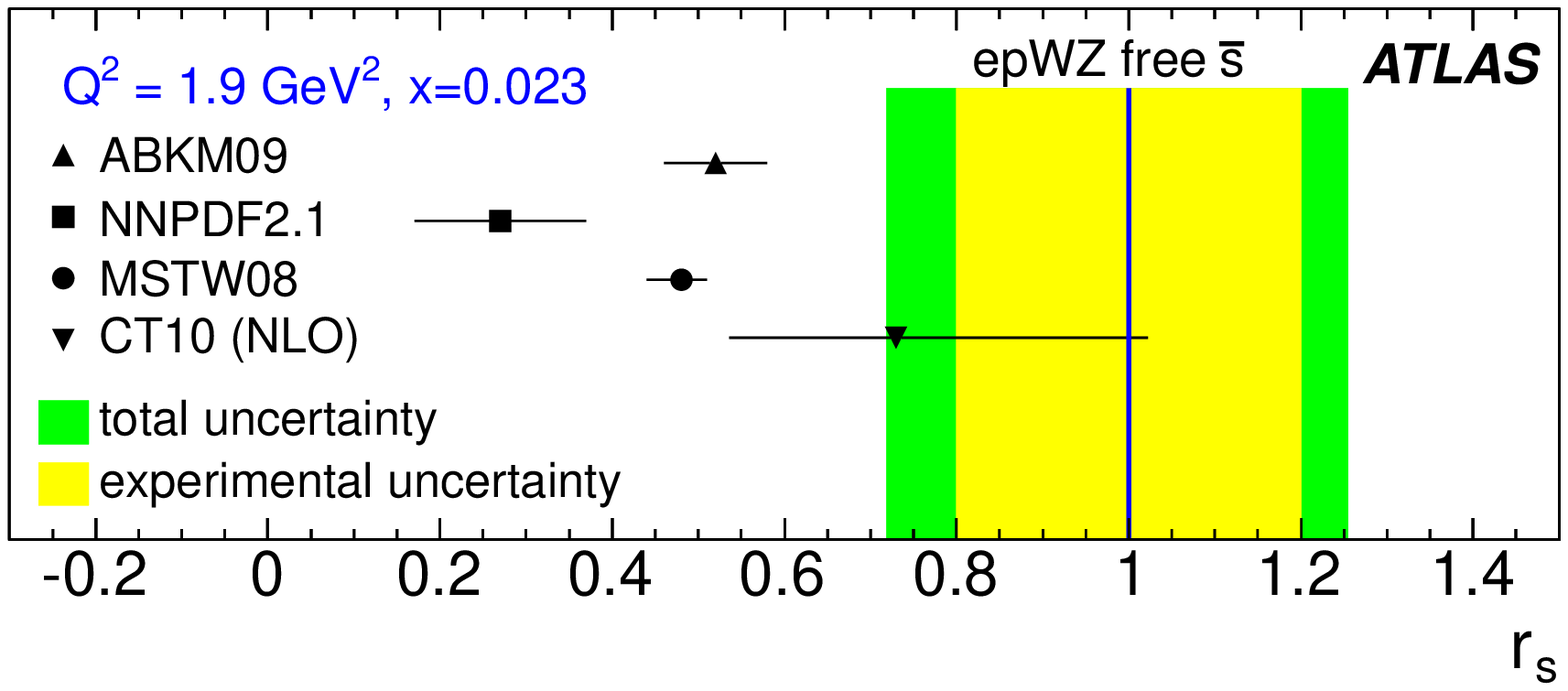,width=0.45\textwidth}}
  \caption{Left: ATLAS data on the $Z$ rapidity distribution from combined 
electron and muon data. The data are compared to two PDF fits, one with fixed 
suppressed strangeness and the other with free unsuppressed strangeness 
as described in the text. 
Right: the ratio of strange to down quark densities in the sea, at 
$x=0.023$, $Q^2=1.9$GeV$^2$, from the preferred epWZ free $\bar{s}$ fit and 
from other PDF determinations.}
  \label{fig:atlasWZ}
\end{figure}

ATLAS have presented $W$-lepton and $Z$ differential cross sections. 
as a function of (pseudo-)rapidity and as a function of 
$p_t$ from $36$pb$^{-1}$ of 2010 data. The $Z$ rapidity distribution from 
combined electron and muon decay channels is shown in Fig.~\ref{fig:atlasWZ}.
Data is also available in the $\tau$ decay channels and measurements of 
$W$ and $\tau$ polarisation have been made~\cite{Sauvan}.
The $W,Z$ rapidity distributions are supplied with full information on 
correlated systematic uncertainties and this allows them to have impact on PDF 
fits. In particular, in Fig.~\ref{fig:atlasWZ} the $Z$ rapidity 
data are presented compared 
to two PDF fits done using these ATLAS $W^{\pm}$ and $Z$ data together with
 the 
HERA DIS data. The fit labelled 'epWZ fixed $\bar{s}$' has the strange quark 
density suppressed and fixed to $50\%$ of the down sea quark density at 
the starting scale of PDF evolution $Q^2_0\sim 2~$GeV$^2$
(as suggested by neutrino di-muon data). The fit labelled 
'epWZ free $\bar{s}$' 
allows the strange quark distribution freedom in normalisation and shape, 
with the result that ATLAS data clearly prefer unsuppressed strangeness for 
$x\sim 0.01$~\cite{klein}. Fig.~\ref{fig:atlasWZ} compares the ratio of the 
strange to down sea quark densities at $x=0.023$, $Q^2_0=1.9~$GeV$^2$, from the
preferred epWZ free $\bar{s}$ fit to that of other PDF determinations.

Measurements of $Z$ or $W$ boson production associated with heavy quarks in the
final state can provide important information about heavy quark densities in PDFs. 
The process $pp \to W + c +X$ and the cross section ratios $R_{+/-}=\sigma(W^{+}c+X)/\sigma(W^{-}c+X)$ 
and $R_{c}=\sigma(W+c+X/\sigma(W+jets+X)$ measured at CMS~\cite{CMSheavy} provide information about 
the strange and anti-strange quark parton density functions of the proton. 
In these measurements muonic decays of the $W$-boson and lifetime
tagging techniques are used to extract the charm fraction in $W+jet$ events.
The measured ratios are: $R_{+/-}=0.92\pm 0.19(stat.)\pm 0.04(syst.)$ and 
$R_{c}=0.143\pm 0.015(stat.)\pm 0.024(syst.)$. 
Currently the 20$\%$ total uncertainty of these results limits their 
constraining 
power but higher statistics samples will significantly
improve the sensitivity to the strangeness content of the PDFs.
Results on the production of b jets in association with $Z/\gamma^*$ were also 
presented by CMS collaboration~\cite{CMSheavy}.

\begin{figure}[htb]
  \centering
\centerline{
\epsfig{figure=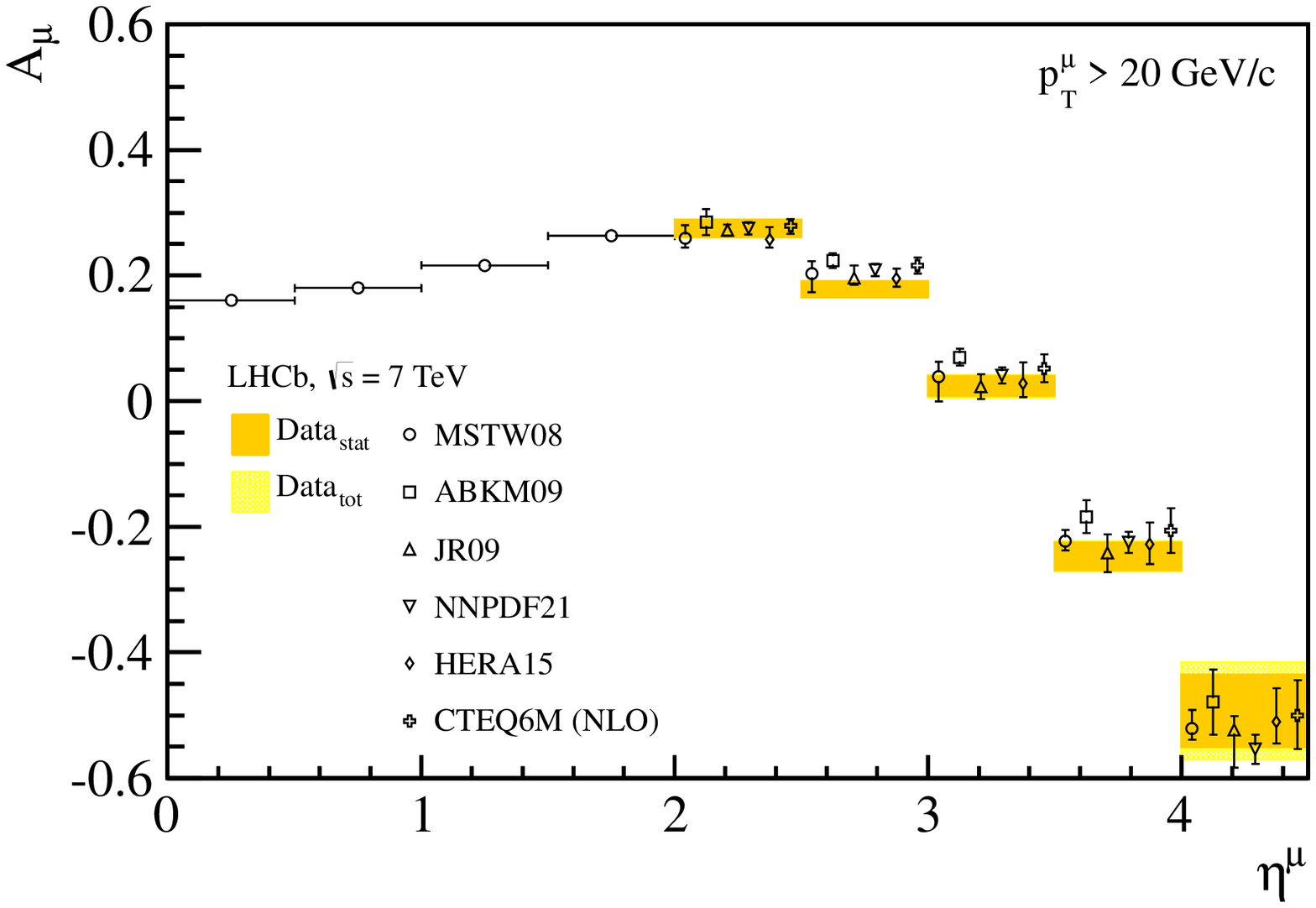,width=0.45\textwidth}
\epsfig{figure=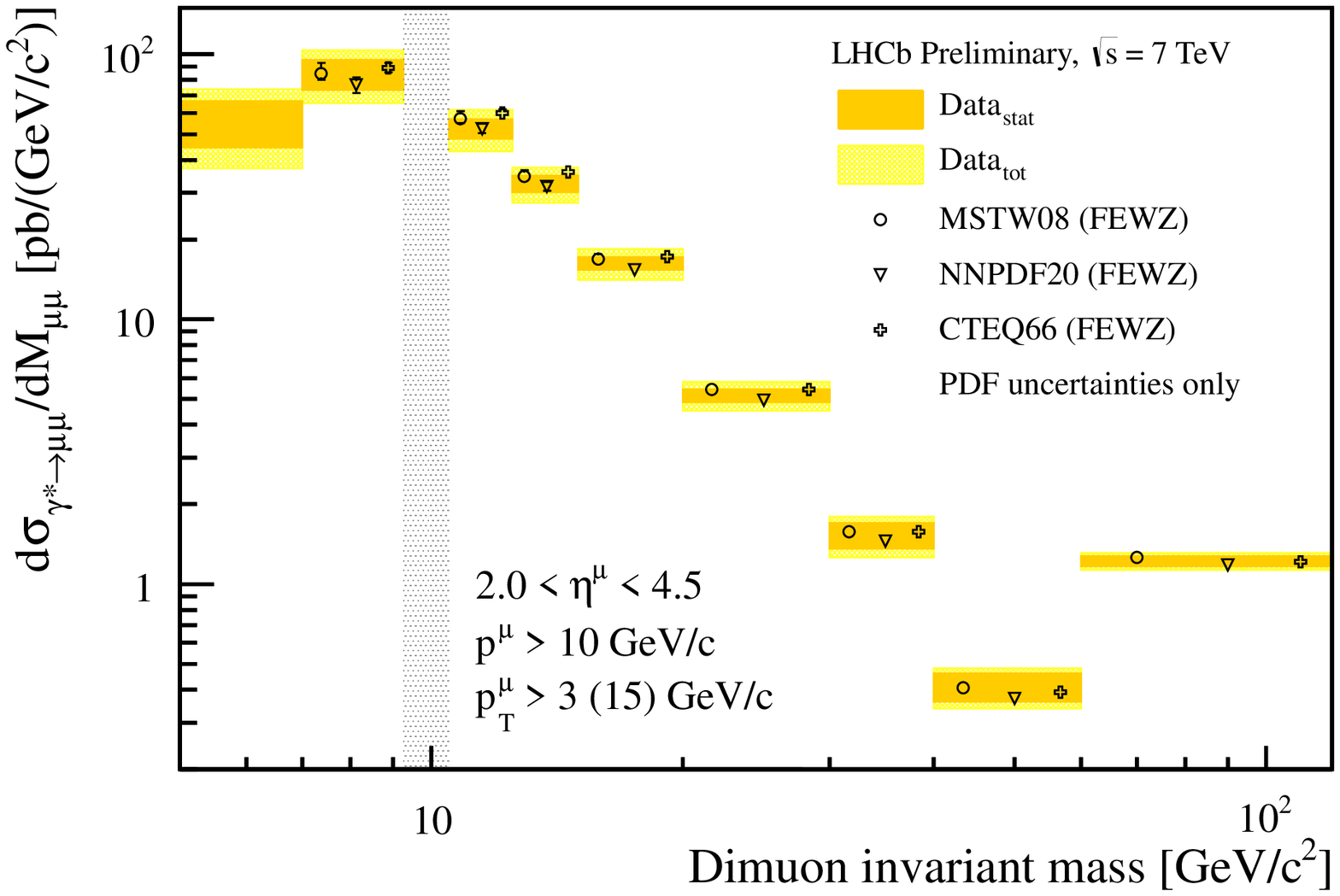,width=0.45\textwidth}}
  \caption{Left: Lepton charge asymmetry $A_{\mu}$ 
in bins of muon pseudorapidity where dark shaded bands correspond 
  to the statistical (dark shaded/orange) and systematic (light hatched/yellow)
experimental uncertainties. These are compared to NNLO and NLO predictions 
with different PDF sets shown by symbols with error bars
  where the PDF and the theoretical uncertainties are added in quadrature
Right: Differential DY cross section as a function of the dimuon invariant mass. 
      Measurements  
          are compared to theoretical predictions calculated with FEWZ using different PDF sets. }
  \label{fig:lhcbasym}
\end{figure}

The LHCb experiment has also performed measurements of $W$ and $Z$ production
using final states containing muons, electrons and tau 
leptons~\cite{LHCbZWtalk}. The LHCb data provide unique constraints on both low-$x$ and high-$x$ PDFs beacuse of the 
high pseudorapidity region ($\eta > 2.5$) in which the $W$ and $Z$ boson cross sections 
are measured. 
Differential cross sections, $W$ and $Z$ cross section ratios and the lepton charge asymmetry 
are measured in this kinematic region. Fig.~\ref{fig:lhcbasym} shows the $W$-lepton asymmetry at high rapidity.
The results presented use mainly 2010 data such that they are still limited 
by statistical precision and the uncertainty on the luminosity. Precision is expected
to improve significantly with the full 2011 dataset.

LHCb have also performed the first low mass Drell-Yan cross section measurements~\cite{LHCbDY} 
for which theoretical uncertainties (particularly scale uncertainties)
 and PDF uncertainties are larger compare to $W$ and $Z$
measurements. Differential DY cross sections are measured as a function of 
dimuon invariant mass (starting from $5~$GeV, see Fig~\ref{fig:lhcbasym}) 
and as a function of rapidity. 
This preliminary result uses background template technique~\cite{LHCbDY}
and currently is limited in statistical and systematic precision.
Significantly smaller uncertainties in the measurement reaching even lower masses are expected 
with 2011 data set where luminosity is about 30 times higher.

\begin{figure}[htb]
  \centering
\centerline{
\epsfig{figure=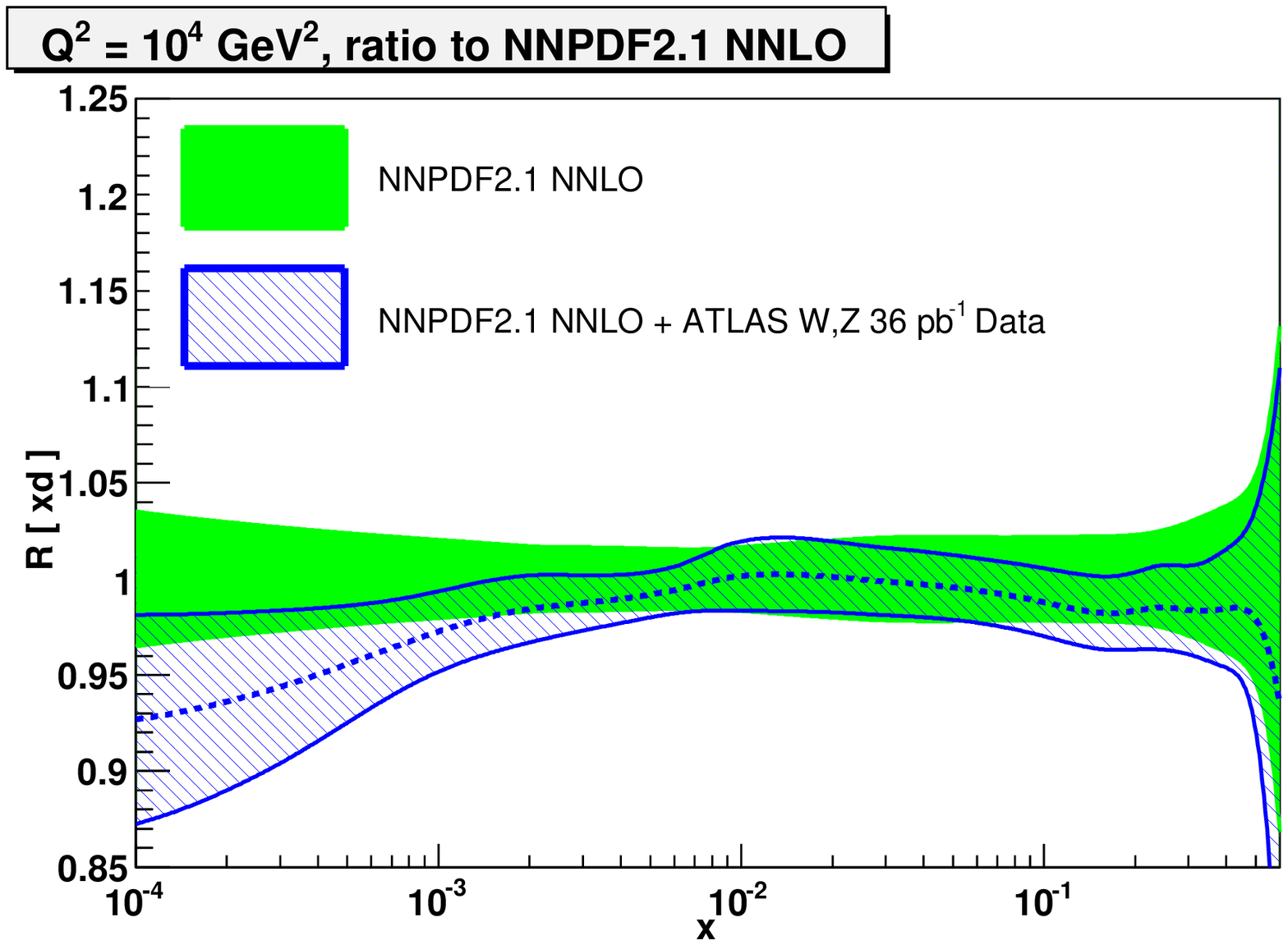,width=0.45\textwidth}
\epsfig{figure=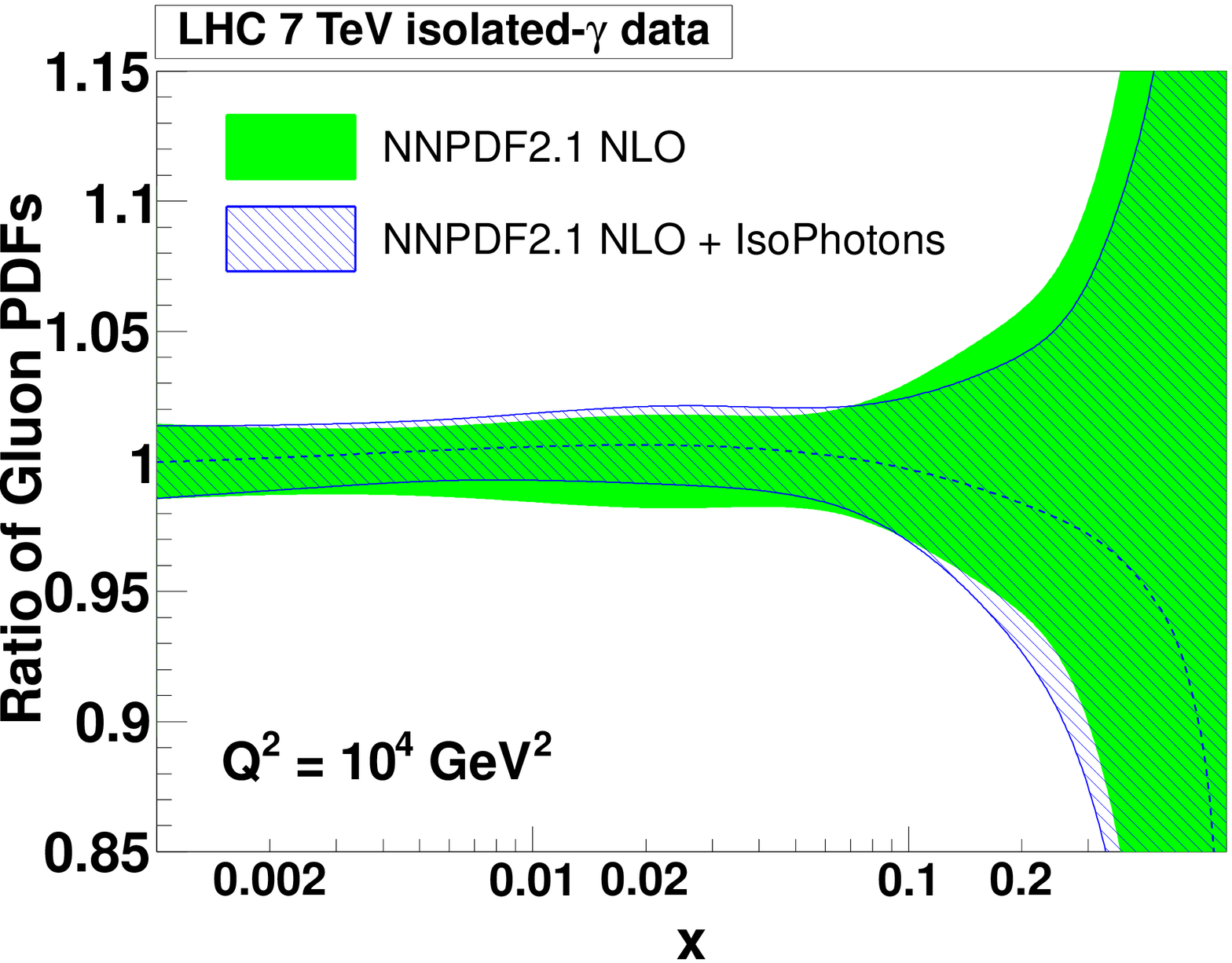,width=0.45\textwidth}}
  \caption{
Left: The ratio of the NNPDF2.1 down quark PDF before and 
after reweighting to include the ATLAS $W,Z$ data, including uncertainties.
Right: The ratio of the NNPDF2.1 gluon PDFs before and after 
reweighting to include prompt photon data, including uncertainties}
\label{fig:nnpdfdde}
\end{figure}

The NNPDF collaboration have examined the impact of the LHC data on PDF fits 
but the early LHC data which was included in the 
NNPDF2.2 set are now superceded. At this meeting a preliminary examination of 
the impact of more recent LHC data on the NNPDF fit was 
presented~\cite{ubiali}. This includes the ATLAS 2010 $W$-lepton
 and $Z$ rapidity 
distributions, the higher luminosity CMS $W$-lepton asymmetry measurements, 
the LHCb $Z$ and $W$-lepton high rapidity distributions, and the ATLAS and CMS 
inclusive jet measurements. The impact of these data has been evaluated by the 
technique of PDF reweighting. Of all these data sets the most discriminating 
is the ATLAS $W^{\pm}$-lepton and $Z$ data. 
Their discriminating power is such that 
of $1000$ initial Monte-Carlo replicas form NNPDF2.1 only 16 survive the 
reweighting procedure. Fig.~\ref{fig:nnpdfdde} shows the impact of these data 
on the down quark PDF shape and uncertainty.
Such a large change necessitates a fresh PDF fit which is underway.

\begin{figure}[htb]
  \centering
  \includegraphics[width=0.68\textwidth]{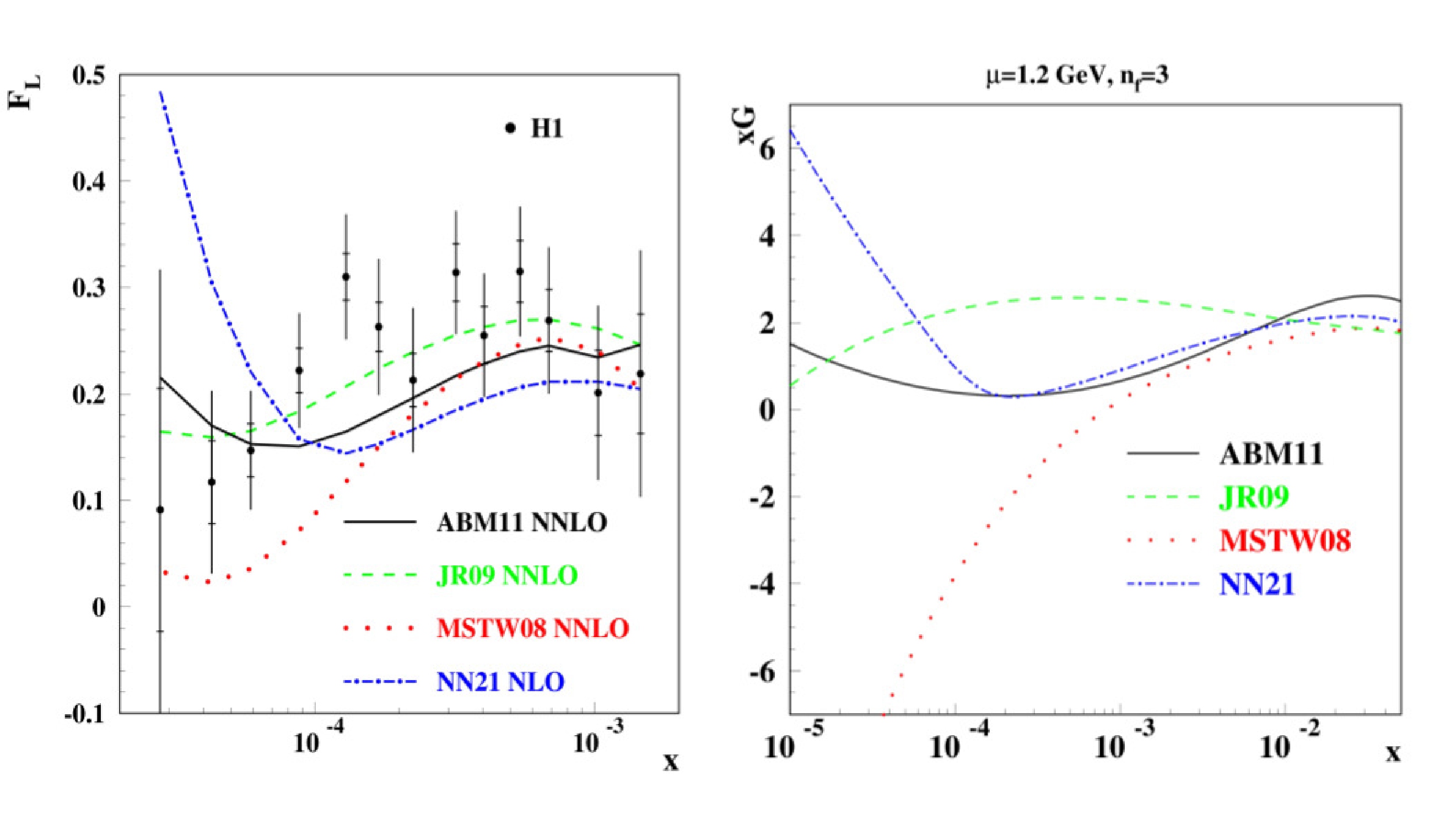}
  \caption{Discriminating power of current $F_L$ data between different gluons at small $x$ \cite{Alekhin}.}
  \label{FL}
\end{figure}

The latest ABM11 PDF analysis~\cite{Alekhin} uses an improved treatment of 
heavy quark electroproduction within the fixed-flavor-number scheme at NNLO. 
This includes the $\overline{\rm MS}$ scheme for heavy quark masses. 
They find good agreement with the latest (combined) HERA charged current data, 
in particular 
no indication of \emph{large logs} up to high $Q^2$. 
The discriminating power of current $F_L$ measurements in the small $x$ region 
is shown in Fig.~\ref{FL}. A similar situation is pointed out for dimuon 
production data. The ABM determination of $\alpha_s(M_Z^2)$ is lower than 
that reported by MSTW and/or NNPDF. It is often suggested that this is because 
jet data are not included in the ABM fit. However ABM find good 
agreement with jet data. They suggest that the higher values found by other 
groups could orginate in higher twist effects in the fixed target DIS data. 
Such power corrections are explicitly included in the ABM formalism.

The ongoing update of the NNLO \emph{dynamical} parton distributions was 
presented\cite{Jimenez-Delgado}. This includes several improvements in the
 framework, e.g. a more careful determination of the strange-quark input 
parton distributions, a complete treatment of the correlations of systematic 
uncertainties of the data, and an improved treatment for heavy quark 
electroproduction from~\cite{Alekhin}; a wealth of deuteron DIS data which 
were not included in the JR09 analysis is also a major update. 
Nevertheless, there are only moderated changes of $10\%$ or less with respect 
to JR09. In particular, the value of $\alpha_s(M_Z^2)$ is determined to be 
$\sim0.113 \to 0.114$, depending on the value of the input scale used; 
the difference between these values is regarded as a genuine uncertainty 
of the determination~\cite{qsrole}.

The CTEQ collaboration has also presented their NNLO results \cite{Nadolsky}. 
There are two different sets at NNLO, one based on pre-LHC data only 
(CT10) and one which will include LHC data on $W$ and $Z$ rapidity 
distributions (CT12). There are ongoing investigations on the 
flavor structure of the quark sea at $x < 10^{-2}$; in particular the 
possibility of $\bar{d}\neq\bar{u}$ as $x \rightarrow 0$, as well as the 
relative size of the strange-quark distribution. Benchmark cross-sections at 
NNLO are presented for the first time.

The CTEQ-JLab Collaboration (CJ)~\cite{Keppel} 
focus on the determination of PDFs in the 
large--$x$ region, with the aim of using the unique capabilities of the CEBAF 
accelerator to measure small cross sections at extreme kinematics to reduce 
the large PDFs uncertainties at large $x$. This requires a careful account of 
corrections which are sometimes suppressed by kinematic cuts in other analyses,
 e.g. higher-twist terms, target mass corrections, nuclear corrections, etc. 
The current focus is on different nuclear corrections in NLO predictions 
for parton 
distributions and structure functions.

A very careful re-analysis of non-singlet world data on unpolarized structure 
functions in the valence region was presented in ref.~\cite{Bluemlein}.  
NLO, NNLO and even approximate N$^3$LO expressions are used 
in the valence region to 
precisely extract the value of the strong coupling constant $\alpha_s(M_Z^ 2)$ 
and the higher twist contributions. The contributions of 
twist $\tau = 3$ and higher to the polarized structure functions $g_1(x,Q^2)$ 
and $g_2(x,Q^2)$ are also investigated. 
At NNLO the value $\alpha_s(M_Z^ 2) = 0.1132 \pm 0.0022$ is obtained
in good agreement with other determinations \cite{Alekhin,Jimenez-Delgado}.

An alternative approach to PDFs is based on the statistical model in which the 
nucleon is regarded as a gas of massless partons in equilibrium at a given 
temperature in a volume of finite size. Only 9 free parameters are adjusted in 
order to completely determine the unpolarized and polarized (helicity) 
distributions. This model produces a reasonable qualitative description, 
even for some data which were not included in the fits. 
Updates to include more recent data in the framework are under consideration.

Before the year 2000 prompt photon data had been used as a means of 
contraining the gluon distribution. However, it was dropped due to 
discrepancies between some of the fixed target data sets and theoretical 
predictions. There has recently been a 
suggestion to re-instate prompt photon data~\cite{denterria} using only high 
hadron collider data, which agree well with predictions made using JETPHOX. 
The impact of these data on the NNPDF2.1 fit has been evalaued by PDF 
reweighting and the data have some impact on the gluon distribution at 
$x\sim 0.01$, as illustrated in Fig.~\ref{fig:nnpdfdde}

The nCTEQ group reported \cite{Kovarik} 
difficulties in describing simultaneously the commonly used charged-lepton 
DIS and Drell--Yan dilepton production off nuclear targets, and data on 
neutrino-nuclei scattering. The source of the problem involves 
the data/error estimates of the NuTeV neutrino data for which the
 correlations of systematic errors need to be taken into account.
This is in contrast to the results presented in \cite{Sassot}, where an 
consistent picture of universal nuclear modification factors was reported. 
They find it possible to describe the main features of all nuclear data, 
including neutrino DIS as well as inclusive pion production, without finding 
any significant tension among the different data sets.

 An analysis of HERA data on the proton structure 
function $F_2$ in the low--$x$ regime using BFKL evolution was presented in 
ref.~\cite{Salas}. A NLL framework 
which includes running coupling effects and makes use of collinear improved 
resummation is used to achieve a good description of the data and to 
study theoretical uncertainties.

The process $W/Z/DY$ + jet, where the boson is produced in the forward 
direction of one of the colliding protons and the jet is produced in the 
forward direction of the second proton, has been proposed for searches of 
evidence for BFKL evolution~\cite{Hentschinski}. 
First numerical results for a number of observables which allow the 
isolation of BFKL effects were presented.

Another contribution within the BFKL framework was the calculation of 
the next-to-leading order photon impact factor for small--$x$ 
DIS~\cite{Chirilli}. An analytic expression in momentum space is derived 
using the operator product expansion in Wilson lines.

The CCFM unintegrated PDF for the gluon has been determined~\cite{Jung} 
using the combined HERA data. For a good 
description of HERA data a calculation of the gluon splitting 
function including non-singular 
terms, imposition of kinematic constraints and an NLO treatment of $\alpha_s$ 
are all necessary. The analysis has been supplemented with an error 
estimation which allows the study of the uPDF uncertainty for processes at 
HERA and LHC.

Extensions of the CCFM evolution equations have been studied~\cite{Krzysztof} 
with the aim of addressing effects like parton 
saturation in final states at the LHC. The question of how to combine the 
physics 
of the BK and CCFM evolution equations has been investigated and a possible 
non-linear 
extension of the CCFM equation has been obtained, 
as suggested by an exclusive form of the BK equation.

General bounds on the ratio of structure functions $F_L/F_2$ have been derived 
within the dipole picture~\cite{Ewerz}, which are valid for any dipole 
cross-section and sharpened by including information on the charm structure 
function $F_2^c$. The bounds are respected by 
the data within the experimental errors, although for 
$3.5 {\rm \,GeV}^2 \leq Q^2 \leq 20 {\rm \, GeV}^2$ the central values of the 
data are close to (and in some cases even above) the bounds, and 
thus put some strain on the validity of the dipole model.

 An effort to understand the small-$x$ 
behavior of the structure function $F_2$ on general grounds 
was presented~\cite{Nachtmann}. It is shown 
that studies in field theory indicate that this behavior may be described 
as a \emph{critical} phenomenon. Under the assumption of a simple power law 
behavior of the matrix elements in the scattering region one can derive an 
expression for $F_2$ which depends on some critical indices which should be 
calculable using lattice methods. A phenomenological extraction of these 
coefficients lead to values similar to those obtained with a two pomeron fit.

\section{Conclusion}
The Structure Functions session at DIS2012 was very lively with many new 
experimental results, especially from the LHC on Drell-Yan production, 
including $W$ and $Z$, and on jet production, including jets with heavy 
flavours. There has been substantial progress in the 
development of tools for parton fitting. The determinations of parton 
distribution functions from different groups still give rise to some 
controversy but there is progress in understanding the differences and progress
in refining the calculations which go into these analyses. All groups 
now present PDFs up to NNLO in the DGLAP formalism. 
The analysis of nuclear PDFs is coming of age.
There has also been 
progress in calculations which extend this formalism into the BFKL regime and 
the high-density regime. 
  
{\raggedright\section*{Acknowledgements}

We thank the organisers for the invitation to convene this session and we 
thank all the speakers in our session for their contributions and the lively discussions.
This research is supported in part by the Swiss National Science Foundation (SNF) under contract 200020-138206. Authored by Jefferson Science Associates, LLC under U.S. DOE Contract No. DE-AC05-06OR23177. The U.S. Government retains a non-exclusive, paid-up, irrevocable, world-wide license to publish or reproduce this manuscript for U.S. Government purposes.

\section{Bibliography}


{\raggedright
\begin{footnotesize}

\end{footnotesize}
}


\end{document}